\documentclass{tlp}
\usepackage{latexsym}
\usepackage{amssymb}
\usepackage{aopmath}
\usepackage{pslatex}
\title[Improving Prolog Programs: Refactoring for Prolog]{Improving Prolog Programs: Refactoring for Prolog}

\author[Alexander Serebrenik, Tom Schrijvers and Bart Demoen]
         {ALEXANDER SEREBRENIK\\
         Laboratory of Quality of Software (LaQuSo), T.U. Eindhoven\\
         HG 5.91, Den Dolech 2, P.O. Box 513, 5600 MB  Eindhoven,
         The Netherlands\\
         \email{A.Serebrenik@tue.nl}
         \and
         TOM SCHRIJVERS\thanks{Research Assistant of the Fund for Scientific Research-Flanders (Belgium)(F.W.O.-Vlaanderen)}\\
         Department of Computer Science, K.U. Leuven\\
	 Celestijnenlaan 200A, B-3001, Heverlee,
	 Belgium\\
         \email{Tom.Schrijvers@cs.kuleuven.ac.be}
         \and 
         BART DEMOEN \\
         Department of Computer Science, K.U. Leuven\\
	 Celestijnenlaan 200A, B-3001, Heverlee,
	 Belgium\\
         \email{Bart.Demoen@cs.kuleuven.ac.be}}

\usepackage{epsfig}
\usepackage{wrapfig}
\usepackage{fancyvrb}
\usepackage{subfigure}
\usepackage{latexsym}
\usepackage{pslatex}
\usepackage{amssymb}

\newcounter{xlisting}[section]
\renewcommand{\thexlisting}{\thesection.\arabic{xlisting}}
\newcommand{\listingcaption}[1]{{\refstepcounter{xlisting}Listing \thexlisting\ - #1}}

\newcounter{rfcounter}[subsection]

\renewcommand{\therfcounter}{\thesubsection.\arabic{rfcounter}}

\newenvironment{refactoring}[1]{\vspace{2mm}\addtocounter{rfcounter}{1}\textbf{\therfcounter\ #1}}{\vspace{2mm}}
\submitted{24 December 2004}
\revised{28 August 2005, 20 September 2006}
\accepted{12 February 2007}

\begin{document}


\newcommand{\vipress}{{\tt ViPReSS} }
\newcommand{\vipressnospace}{{\tt ViPReSS}}
\newcommand{\btw}{{\sf BTW} }
\newcommand{\btwnospace}{{\sf BTW}}
\newcommand{\codeframesep}{1mm}
\newcommand{\eat}[1]{}


\maketitle

\begin{abstract}

{\em Refactoring} is an established technique from the object-oriented (OO)
programming community to restructure code: it aims at improving software
readability, maintainability and extensibility. Although refactoring is not
tied to the OO-paradigm in particular, its ideas have not been applied to
Logic Programming until now.

This paper applies the ideas of refactoring to Prolog programs.  A catalogue
is presented listing refactorings classified according to scope. Some of
the refactorings have been adapted from the OO-paradigm, while others have been
specifically designed for Prolog. The discrepancy between intended
and operational semantics in Prolog is also addressed by some of the refactorings.

In addition, \vipressnospace, a semi-automatic refactoring browser,
is discussed and the experience with applying \vipress to a large Prolog
legacy system is reported. The main conclusion is that refactoring is both
a viable technique in Prolog and a rather desirable one.

\end{abstract}


\section{Introduction}\label{sec:refactoring}

Maintaining and adapting software takes up a substantial part of the entire
programming effort, both in time and money. Erlikh~\shortcite{Erlikh}
and  Moad~\shortcite{Moad} both report on the proportion of maintenance
costs exceeding 90\% of the budget.  About 75\% of these costs are spent on
providing enhancements (in the form of adaptive or perfective maintenance)
\cite{Nosek:Palvia,vanVliet}. 

Before providing enhancements, it is recommended to improve the design
of the software in a preliminary step. This methodology, called
{\em refactoring}, emerged from a number of pioneer results in the
OO-community~\cite{Fowler:et:al,Opdyke:PhD,Roberts:Brant:Johnson} and
recently came to prominence for functional~\cite{Li:Reinke:Thompson} and
procedural~\cite{Garrido:Johnson} languages.

Refactoring is a disciplined technique for restructuring an existing body of
code, altering its internal structure without changing its external behavior.
Its heart is a series of small source-to-source program transformations,
called {\em refactorings}, that change program structure and organization,
but not program functionality. The major aim of refactoring is to improve
readability, maintainability and extensibility of the existing software.

While performance improvement is not considered as a crucial issue for
refactoring, it can be noted that well-structured software is more amenable to
performance tuning. We also observe that certain techniques that were developed
in the context of program optimization, such as dead-code elimination and
redundant argument filtering, can improve program organization and, hence,
can be used as refactoring techniques.

In this paper we study refactoring techniques for Prolog. Our goals are
threefold. Firstly, we want to show that refactoring is a viable technique
for Prolog and many of the existing techniques developed for refactoring
in general are applicable. Secondly, Prolog-specific refactorings are
possible and the application of some general techniques may be highly
specialized towards Prolog. 
Finally, it should
be clear that refactoring is not only viable for Prolog but also very useful for
the maintenance of Prolog programs.

In order to achieve our goals we present a catalogue of refactoring
techniques for Prolog. The listed refactorings are a mix of general and
Prolog-specific ones. 
Most of the refactorings proposed have been implemented in a prototype 
refactoring browser \vipressnospace. \vipress has been successfully applied for
refactoring a 50,000 lines-long legacy system. 

As completeness of the catalogue is clearly not possible, we aimed to show a
wide range of possibilities for future work on combining the formal techniques
of program analysis and transformation with software engineering.
The formal elaboration of a particular topic may be a
substantial study on its own, as shows the work on detecting duplicate code
by Vanhoof~\shortcite{Vanhoof} that was inspired by a preliminary version
of our work.

\paragraph{Outline of the Paper}
First, Section \ref{sec:process} provides a brief overview of the refactoring process.
Next, 
the use of several refactoring techniques is illustrated on a small example in
Section \ref{sec:example}. Then a catalogue of Prolog 
refactorings is given in Section \ref{sec:catalogue}. 
In Section \ref{sec:vipress} we introduce \vipressnospace, and discuss
its application in a case study. Finally,
in Section \ref{sec:conclusions} we conclude.

\section{The Refactoring Process}\label{sec:process}

The refactoring process consists of applying a number of refactorings, with both localized and global impact, to a
software system.  
The individual significance of
a refactoring may be apparent, but 
often a refactoring seems trivial on
its own and only in conjunction with other refactorings or intended changes
does the usefulness become clear. That is the reason why it is not feasible
to fully automate refactorings. They must be carefully considered in view
of the programmer's intentions.

For this reason the process of applying a single refactoring is to be split
into a number of distinct activities \cite{Mens:Tourwe}. These activities
involve decisions to be made by the programmer.

The first decision is {\em where} the software should
be refactored. Making this decision automatically can be a difficult
task on its own. Several ways to resolve this may be considered. For
instance, one can aim at identifying so called {\em bad smells}, i.e.,
``structures of the code that suggest (sometimes scream for) the possibility of
refactoring''~\cite{Fowler:et:al}. To this end program analysis can
be used. For example, it is common practice while ordering predicate arguments
to start with the input arguments and end with the output arguments. Mode
information can be used to detect when this rule is violated.

Next, one should determine {\em which} refactorings should be applied.
Sometimes, the correspondence between bad smells and refactorings
is clear. For instance, if the predicate arguments are not ordered according
to the ``input first output last'' rule, one can suggest to the user to reorder 
the arguments. This refactoring is further discussed in Section~\ref{section:reorder:arguments}. 
In more complex situations the relation becomes less obvious:
a number of different refactorings are applicable  and the user has to choose between them.
For example, let module \texttt{A} contain a predicate that is mutually recursive
with predicate \texttt{p} from module \texttt{B}, and module 
\texttt{C} contain a predicate that is mutually recursive
with predicate \texttt{q} from module \texttt{B}. This situation can be identified as problematic
since no clear hierarchy can be defined between these modules.
One possible solution would be to merge the three modules (Section~\ref{section:merge:modules}). 
Alternatively, one may try to
first split \texttt{B} into \texttt{B1}, containing \texttt{p}, and \texttt{B2} containing \texttt{q}
such that there are no circular dependencies between \texttt{B1} and \texttt{B2} (Section~\ref{section:split:module}). 
If this split is possible,
\texttt{A} could be merged with \texttt{B1}, and \texttt{C} with \texttt{B2} 
(Section~\ref{section:merge:modules}). 
Automatic refactoring tools, so called {\em
refactoring browsers}, can be expected to make suggestions on where refactoring
transformations should be applied. These suggestions can then be either
confirmed or rejected by the programmer.

By definition, refactorings should preserve the software's functionality.
Hence, the next step consists of {\em ensuring} that the behavior is indeed
preserved. This step, of course, depends on the definition of behavior.
In the case of logic programming, behavior comprises computed answers 
semantics,
termination, and side effects such as input/output. 
It should be observed that particular application domains might
require extending the notion of behavior to include such concepts as
efficiency or memory use. Moreover, in order for some refactorings to be 
applicable certain preconditions should hold, like absence of user-defined
meta-predicates for dead-code elimination discussed in Section~\ref{section:remove:dead:code}. 
Sometimes verification of the
preconditions cannot be done automatically, but must be delegated to the user. 

Subsequently, {\em the chosen transformation is applied}. This step
might also require 
user input. Consider for example a refactoring that renames a predicate:
while automatic tools can hardly be expected to guess the new predicate name,
they should be able to detect all program points affected by the change. This 
refactoring is further studied in Section~\ref{section:rename:predicate}.

Finally, the {\em consistency} between the refactored program code and 
other related artifacts should be maintained. By artifacts we understand 
among others software documentation, specifications and test descriptions.
The ability to perform this task automatically strongly depends on the 
formalisms used to express the corresponding artifacts. 
For instance, documentation generators such as {\em lpdoc}~\cite{Hermenegildo}
make it possible to keep the documentation consistent automatically, whereas
ad hoc unstructured comments are much harder to update automatically. Ensuring
consistency is considered as future work.


\section{Detailed Prolog Refactoring Example}\label{sec:example}
\label{section:example}

We illustrate some of the techniques proposed by a detailed refactoring example.
Consider the following code fragment from O'Keefe's ``The Craft of
Prolog'' \shortcite{OKeefe}, p. 195. It describes three operations on a {\em reader}
data structure used to sequentially read terms from a file. The three
operations are \texttt{make\_reader/3}, which initializes the data structure,
\texttt{reader\_done/1}, which checks whether no more terms can be read, and
\texttt{reader\_next/3}, which gets the next term and advances the reader.

\begin{Verbatim}[commandchars=\\\{\},frame=single,fontsize=\small,framesep=\codeframesep,label={[\listingcaption{O'Keefe's original version}]}]
\textbf{make_reader}(File,Stream,State) :-
        open(File,read,Stream),
        read(Stream,Term),
        reader_code(Term,Stream,State).

\textbf{reader_code}(end_of_file,_,end_of_file) :- ! .
\textbf{reader_code}(Term,Stream,read(Term,Stream,Position)) :-
        stream_position(Stream,Position).

\textbf{reader_done}(end_of_file).

\textbf{reader_next}(Term,read(Term,Stream,Pos),State)) :-
        stream_position(Stream,_,Pos),
        read(Stream,Next),
        reader_code(Next,Stream,State).
\end{Verbatim}

We will now apply several refactorings to the above program in order to improve
its readability.

Firstly, we use if-then-else introduction (Section~\ref{section:replace:cut:by:if-then-else}) 
to get rid of the red cut\footnote{As defined in e.g. \cite{OKeefe}: a cut that alters the meaning.} in the \texttt{reader\_code/3} predicate
(modified code is underlined):


\begin{Verbatim}[commandchars=\\\{\},frame=single,fontsize=\small,framesep=\codeframesep,label={[\listingcaption{Replace cut by if-then-else}]}]
\textbf{reader_code}(Term,Stream,State) :-
        \underline{( Term = end_of_file,}
          \underline{State = end_of_file ->}
                true
        \underline{;}
                State = read(Term,Stream,Position),
                stream_position(Stream,Position)
        \underline{)}.
\end{Verbatim}

The result of this automatic transformation reveals two malpractices: the first is
producing output before the commit, something O'Keefe himself disapproves of
in \shortcite{OKeefe}. This malpractice and the ways to resolve it are further investigated 
in~\ref{section:produce:output:after:commit}.
The problem is fixed to:


\begin{Verbatim}[commandchars=\\\{\},frame=single,fontsize=\small,framesep=\codeframesep,label={[\listingcaption{Output after commit}]}]
\textbf{reader_code}(Term,Stream,State) :-
        ( Term = end_of_file ->
                \underline{State = end_of_file}
        ;
                State = read(Term,Stream,Position),
                stream_position(Stream,Position)
        ).
\end{Verbatim}

The second malpractice is a unification in the condition
of the if-then-else where an equality test is meant.
Consider the case that the \texttt{Term} argument is a variable. Then
the binding of \texttt{Term} to the atom \texttt{end\_of\_file} is certainly unwanted behavior.
The transformation in question is discussed in Section~\ref{section:replace:unification:by:inequality:test}.
The following code does not exhibit the problematic behavior: 


\begin{Verbatim}[commandchars=\\\{\},frame=single,fontsize=\small,framesep=\codeframesep,label={[\listingcaption{Equality test}]}]
\textbf{reader_code}(Term,Stream,State) :-
        ( \underline{Term == end_of_file} ->
                State = end_of_file
        ;
                State = read(Term,Stream,Position),
                stream_position(Stream,Position)
        ).
\end{Verbatim}

Next, we notice that the conjunction \texttt{read/2, reader\_code/3}
occurs twice. By applying predicate extraction (Section~\ref{section:extract:predicate:locally}) 
of this common sequence, we get:


\begin{Verbatim}[commandchars=\\\{\},frame=single,fontsize=\small,framesep=\codeframesep,label={[\listingcaption{Predicate extraction}]}]
\textbf{make_reader}(File,Stream,State) :-
        open(File,read,Stream),
        \underline{read_next_state(Stream,State)}.

\textbf{reader_next}(Term,read(Term,Stream,Pos),State)) :-
        stream_position(Stream,_,Pos),
        \underline{read_next_state(Stream,State)}.

\underline{\textbf{read_next_state}(Stream,State) :-}
        \underline{read(Stream,Term),}
        \underline{reader_code(Term,Stream,State).}
\end{Verbatim}

Next we put the input argument first
and the output arguments last (Section~\ref{section:reorder:arguments} below), 
a principle also advocated in \cite{OKeefe}:


\begin{Verbatim}[commandchars=\\\{\},frame=single,fontsize=\small,framesep=\codeframesep,label={[\listingcaption{\label{lst:reorder_arguments}Argument reordering}]}]
\textbf{reader_next}\underline{(read(Term,Stream,Pos),Term,State)} :-
        stream_position(Stream,_,Pos),
        read_next_code(Stream,State).
\end{Verbatim}

Finally, note that the naming of the two builtins
\texttt{stream\_position/[2,3]} may be confusing to the user. It is easier to
distinguish between their functionality based on predicate name than based on
arity. We introduce the less confusing names \texttt{get\_stream\_position/2}
and \texttt{set\_stream\_position/3} respectively. 
In addition, we provide a more consistent naming for \texttt{make\_reader},
more in line with the other two predicates in the interface. The importance of consistent naming conventions
is also stressed in \cite{OKeefe}.

Note that direct renaming of built-ins such as \texttt{stream\_position}
is not possible, but a similar effect can be achieved by
extracting the built-in into a new predicate with the desired name.
Extracting a predicate and renaming predicates
are considered in Sections~\ref{section:extract:predicate:locally} and 
~\ref{section:rename:predicate}, respectively.

In order to avoid confusion between a built-in predicate 
\texttt{read} and a functor \texttt{read} we rename the latter functor to 
\texttt{reader}.



\begin{Verbatim}[commandchars=\\\{\},frame=single,fontsize=\small,framesep=\codeframesep,label={[\listingcaption{\label{lst:rename_functor}Renaming}]}]
\underline{\textbf{reader_init}}(File,Stream,State) :-
        open(File,read,Stream),
        reader_next_state(Stream,State).

\textbf{reader_next}(\underline{reader}(Term,Stream,Pos),Term,State)) :-
        \underline{set_stream_position}(Stream,Pos),
        reader_next_state(Stream,State).

\textbf{reader_done}(end_of_file).

\textbf{reader_next_state}(Stream,State) :-
        read(Stream,Term),
        build_reader_state(Term,Stream,State).

\textbf{build_reader_state}(Term,Stream,State) :-
        ( Term == end_of_file ->
                State = end_of_file
        ;
                State = \underline{reader}(Term,Stream,Position),
                \underline{get_stream_position}(Stream,Position)
        ).

\underline{\textbf{set_stream_position}(Stream,Position) :-}
        \underline{ stream_position(Stream,_,Position).}
\underline{\textbf{get_stream_position}(Stream,Position) :-}
        \underline{ stream_position(Stream,Position).}
\end{Verbatim}

This example demonstrates how the code readability can be ameliorated by performing
a series of relatively simple transformation steps. We have seen that some of these steps
required user's input. Clearly the changes can be performed manually. 
However, refactoring browsers should be able to guarantee consistency, correctness
and furthermore can automatically single out opportunities for refactoring.

Techniques applied above are well-suited for local code improvement, i.e., 
the objects modified are predicates and clauses. In the next section
we also consider techniques for global code restructuring such as
duplicate predicates removal (Section~\ref{section:remove:duplicate:predicates}).

\section{A Catalogue of Prolog refactorings}\label{sec:catalogue}

In this section we present the refactorings that we have found to be useful
for Prolog programs. 
The considered Prolog programs are
not limited to pure logic programs, but may contain various built-ins such as
those defined in the ISO standard \shortcite{ISO13211-1}. The only exception are 
higher-order constructs that are not dealt with automatically, but manually.
This is done due to the fact that higher order constructs such as {\em call} 
make it impossible to decide at the compile-time which predicate is going 
to be called at the corresponding program point during execution. 
Automating the detection and handling of higher-order predicates is an
important part of future work.

The refactorings in this catalogue are grouped by their scope. The scope expresses
the user-selected target of a particular refactoring. Hence, refactoring starts
by choosing an object in the specified scope. For instance, {\em split module}
(Section~\ref{section:split:module}) starts with selecting a module. 
Then
the object is transformed. 
For us, this means that the module is split. Finally, the changes propagate to
the affected code outside the selected scope. The latter might happen when
there is a dependency outside the scope. This corresponds
to updating import declarations in other modules of the system.

For Prolog programs we distinguish the following four scopes, based on the
code units of Prolog: 
{\em system} scope (Section \ref{sub:system}),
{\em module} scope (Section \ref{sub:module}),
{\em predicate} scope (Section \ref{sub:predicate}) and
{\em clause} scope (Section \ref{sub:clause}).

As a starting point for this catalogue we used Fowler's~\shortcite{Fowler:catalogue}
for object-oriented languages. We selected those with clear Prolog counterparts, extended
the list with Prolog-specific transformations and some well-known program transformations, such as
dead code elimination.

In the current technical note we only include a short summary of the refactorings here
and refer to the companion technical report \cite{techrep}. 
This report contains
the full catalogue with detailed description of the refactorings, examples,
preconditions and automatization techniques.

\subsection{System Scope Refactorings}\label{sub:system}

The system scope encompasses the entire code base. The user wants to consider
the system as a whole.



\begin{refactoring}{Eliminate explicit module qualification}
\label{section:eliminate:explicit:module:qualification}

In many Prolog systems, such as Quintus~\cite{Quintus:Manual}%
, the module system is non-strict, i.e. the normal visibility 
rules can be overridden by a special construct, called {\em explicit module qualification}
and written as \texttt{m:q}
, 
where \texttt{m} is a module that contains definition of the 
predicate \texttt{q/0}. The refactoring
proposed adds import and export declarations to get rid of these special syntax constructions. 
%
By forcing the code to conform to a strict module system a number of quality characteristics are improved. 
First of all, 
a strict module system better expresses the idea of information hiding, 
which is important for software maintainability and readability~\cite{ParnasCriteria}. Moreover, 
since not all Prolog systems support the above construct, code portability is improved. 

\end{refactoring}


\begin{refactoring}{Extract common code into predicates}
\label{section:extract:common:code:into:predicates}

This refactoring looks for 
common functionality across the system
and extracts it into new predicates. The common functionality consists of 
identical subsequences of goals that are called in different predicate bodies, and extracts them
into new predicates. 
%
The overall
readability of the program improves as the affected predicate bodies get
shorter, 
and the calls to the new predicates can be more meaningful than what
they replace. 
Moreover the increased sharing simplifies maintenance 
as now only one copy
needs to be modified.  

The problem of identifying identical subsequences of 
of goals is related to determining longest repeated 
subsequences~\cite{Crow:Smith,Pitkow:Pirolli}.
\end{refactoring}


\begin{refactoring}{Hide predicates}
\label{section:hide:predicates}

This refactoring removes export declarations for predicates that are not
imported in any other module. It simplifies the program by reducing the number of entry
points into modules and hence the intermodule dependencies.


%
\end{refactoring}


\begin{refactoring}{Remove dead code}
\label{section:remove:dead:code}

Dead code is code that can never be executed and therefore can
be safely eliminated without affecting correctness of the execution.
Dead code elimination is sometimes performed in compilers for efficiency
reasons, but it is also useful for developers: dead code clutters the program.
We consider a predicate definition as the unit of dead code.

\end{refactoring}

 
\begin{refactoring}{Remove duplicate predicates}
\label{section:remove:duplicate:predicates}

Predicate duplication or cloning is a well-known problem, 
prominently caused by ``copy \& paste'' and
unawareness of available libraries and exported predicates in
other modules. The main problem with duplication is its bad
maintainability. 
It is up to the user to decide whether to throw away some of the duplicates
and to use one of the remaining definitions instead
or to replace all the duplicate predicates by a new version in a new module.
\end{refactoring}


\begin{refactoring}{Rename functor}
\label{section:rename:functor}

This refactoring renames a term functor across the system. If the functor has
several different meanings and only one should be renamed, it is up to the user to
identify what occurrence corresponds with what meaning. 

\end{refactoring}

\subsection{Module Scope Refactorings}\label{sub:module}

The module scope considers a particular module. Usually a module is
implementing a well-defined functionality and is typically contained in
one file.  



\begin{refactoring}{Merge modules}
\label{section:merge:modules}

Merging several modules into one can be advantageous in case of strong
interdependency of the modules involved. Moreover, merging existing modules
and splitting the resulting module can lead to an improved module
structure.

\end{refactoring}


\begin{refactoring}{Remove dead intra-module code}
\label{section:remove:dead:code:intra-module}

Similar to {\em dead code removal} for an entire system  (see Section
\ref{section:remove:dead:code}), 
this refactoring works at the level of a single module.
It is useful for incomplete systems or library modules with an unknown number
of uses. Recall that determining the liveness of the code requires
knowledge of top-level predicates. In the case of intra-module dead
code elimination, the set of top level predicates is extended with, 
or replaced by, the exported predicates of the module. 

%

\end{refactoring}

 

\begin{refactoring}{Rename module}
\label{section:rename:module}

This refactoring applies when the name of the module no longer corresponds
to the functionality it implements e.g. due to other refactorings. 



%
\end{refactoring}


\begin{refactoring}{Split module}
\label{section:split:module}

The refactoring is useful to split unrelated
parts of a module or make a large module more manageable.  



Moores~\cite{Moores} has shown that the number of user-defined 
predicates correlates with the number of errors 
detected. Based on an empirical study he suggested a threshold of around 
$35\pm 5$ predicates per program. While this is hardly reasonable as a 
requirement for an entire Prolog system, trespassing the threshold should be 
used as a guideline when the Split Module refactoring can be applied.  
%
%
\end{refactoring}

\subsection{Predicate Scope Refactorings}\label{sub:predicate}

The predicate scope targets a single predicate. The code that
depends on the predicate may need updating as well. But this is considered
an implication of the refactoring of which either the user is alerted or
the necessary transformations are performed automatically.



\begin{refactoring}{Add argument}
\label{section:add:argument}

This refactoring should be applied when a callee needs more information 
from its (direct or indirect) caller, which is very common in Prolog program
development. 
Given a variable in the body of the caller and the name 
of the callee, the refactoring browser should propagate
this variable along all possible computation paths from the caller to the
callee. This refactoring is an important preliminary step preceding 
additional functionality integration or efficiency improvement.

\end{refactoring}


\begin{refactoring}{Move predicate}
\label{section:move:predicate}

This refactoring moves a predicate definition from one module 
to another.
It can improve the overall structure of the program by bringing together
interdependent or related predicates,
hence improving both cohesion of each one 
of the modules involved, and coupling of the pair. 
{\em Move predicate} appears often
after predicate extraction, i.e., {\em extract common code} or {\em extract predicate locally},
discussed in Sections~\ref{section:extract:common:code:into:predicates} 
and \ref{section:extract:predicate:locally}, respectively.

\end{refactoring}


\begin{refactoring}{Rename predicate}
\label{section:rename:predicate}

This refactoring can improve
readability and should be applied when the name of a predicate does not reveal
its purpose. 
\end{refactoring}


\begin{refactoring}{Reorder arguments}
\label{section:reorder:arguments}

Our experience suggests that while writing predicate definitions Prolog
programmers tend to begin with the input arguments and to end with the output
arguments. This habit has been identified as a good practice and even
further refined by O'Keefe \shortcite{OKeefe} to more elaborate rules.
Unfortunately, this practice is difficult to maintain when additional arguments
are added later. We observed that failure to confirm to this ``input first
output last'' expectation pattern is experienced as very confusing. 
\end{refactoring}


\begin{refactoring}{Specialize predicate}
\label{section:split:predicate}

By specializing a predicate we mean producing a (number of) more
specific version(s) of a given predicate provided some knowledge on the intended
uses of the predicate. Specialisation can simplify code as well as make a meaningful
distinction between different uses of a predicate.
\end{refactoring}


\begin{refactoring}{Remove redundant arguments}
\label{section:remove:redundant:arguments}

The basic intuition here is that parameters that are no longer used
by a predicate should be dropped. 
It improves readability.



%

 Leuschel and S{\o}rensen~\shortcite{Leuschel:Sorensen} established that the
 redundancy property is undecidable and suggested two techniques to find
 safe and effective approximations: top-down goal-oriented RAF (Redundant 
 Argument Filtering) and bottom-up goal-independent FAR (RAF ``upside-down''). 
 In the context of refactoring FAR is the more useful
 technique, since only FAR deals correctly with exported predicates used 
 in unknown goals. 
\end{refactoring}

\subsection{Clause Scope Refactorings}\label{sub:clause}

The clause scope affects a single clause in a predicate. Usually, this does
not affect any code outside the clause directly. 



\begin{refactoring}{Extract predicate locally}
\label{section:extract:predicate:locally}

This refactoring is similar to the system-scope refactoring with the same name.
However, it does not aim to automatically
discover useful candidates for replacement.
The user is responsible for selecting the subgoal that
should be extracted, in order to improve the readability.

\end{refactoring}


\begin{refactoring}{Invert if-then-else}
\label{section:invert:if-then-else}

The order of ``then'' and ``else'' branches can be important for
code readability. 
%
To enhance readability it might be worthwhile putting the shorter branch as
``then'' and the longer one as ``else''. Alternatively, the negation of
the condition may be more readable because, for example, a double negation can be
eliminated. 
\end{refactoring}

\begin{refactoring}{Replace cut by if-then-else}
\label{section:replace:cut:by:if-then-else}

This technique aims at improving program readability by replacing
cuts (!) by the more declarative if-then-else ({\tt  -> ; }). More detailed
discussion on replacing cut by if-then-else is deferred to {\em Related
work and extensions}.

\end{refactoring}


\begin{refactoring}{Replace unification by (in)equality test}
\label{section:replace:unification:by:inequality:test}

Often full unifications are used instead of equality or other tests.
%
O'Keefe in \shortcite{OKeefe} advocates the importance of steadfast code. Recall, that
steadfast code produces the right answers for all possible modes and inputs. A more moderate
approach is to write code that works for the intended mode only.
%
Unification succeeds in several modes and so does not convey a particular
intended mode. Equality ({\tt ==}, {\tt =:=}) and inequality ({\tt \verb+\==+},
{\tt \verb+=\=+}) checks usually only succeed for one particular mode and
fail or raise an error for other modes. Hence their presence makes it
easier in the code and at runtime to see the intended mode. Moreover, if
only a comparison was intended, then full unification may lead to unwanted
behaviour in unforeseen cases.
\end{refactoring}


\begin{refactoring}{Produce output after commit}
\label{section:produce:output:after:commit}

This refactoring addresses a similar issue as the previous one. Producing
output before the commit (cut) does not properly convey the intended mode
of a predicate. Moreover it may lead to unexpected results when used
in the wrong mode.
\end{refactoring}
 
\section{The \vipress refactoring browser}\label{sec:vipress}

The refactoring techniques presented in Section \ref{sec:catalogue} have been implemented in
the prototype refactoring browser \mbox{\vipressnospace}\footnote{Vi(m) P(rolog)
Re(factoring) (by) S(chrijvers) (and) S(erebrenik)}. 
It has been implemented
on the basis of VIM%
, a popular clone of the well-known VI editor. The text editing facilities of VIM
make it easy to implement techniques like {\em move
predicate} (Section~\ref{section:move:predicate}). 

Most of the refactoring tasks have been implemented as SICStus Prolog
\cite{SICStus:Manual} programs inspecting source files and/or call
graphs. Updates to files have been implemented either directly in the
scripting language of VIM or, when many files need updating
at once, through \texttt{ed} scripts. VIM functions were written to
initiate the refactorings and to get user input.



\vipress has been successfully applied to a large (more than 53 KLOC)
legacy system used at the Computer Science department of the Katholieke
Universiteit Leuven to manage the educational activities. The system, called
\btwnospace, 
has been developed and extended since the
early eighties by more than ten programmers, many of whom are no
longer employed by the department. The implementation has been done in 
MasterProLog~\cite{MasterProLog}, which
is no longer supported. 
 Therefore, preparing the code for migration to a more
 modern Prolog dialect and general structure improvement were essential for
 further evolution of the system.

By using the refactoring techniques we succeeded in obtaining a better
understanding of this real-world system, in improving its structure and
maintainability, and in preparing it for intended changes:
porting it to a state-of-the-art Prolog system and adapting it to
new educational tasks the department is facing as a part of the unified
Bachelor-Master system in Europe.

A preliminary study revealed that many modules were unused. We brought in
an expert to help us identify the bulk of these unused modules, including
out-of-fashion user interfaces and outdated versions of program files. This
reduced the system size to a mere 20,000 lines.


Next, the actual refactoring process was started. As the first phase
we applied system-scope refactorings.
\vipress was used to clean up after the bulk dead code removal:
299 predicates in the remaining modules were identified as dead.
This reduced the size by another 1,500 lines.  Moreover
\vipress discovered 79 pairwise identical predicates. In most of the cases,
identical predicates were moved to new modules used by the original ones. The
previous steps allowed us to improve the overall structure of the program
by reducing the number of files from 294 to 116 with a total of 18,000
lines. Very little time was spent to bring the system into this state. The
experts were sufficiently familiar with the system to identify
obsolete parts. The system-scope refactorings took only a few minutes each.
During this phase most of the work has been done by \vipressnospace,  while the user's 
involvement was limited to choosing a way to deal with duplicate predicates.

The second phase of refactoring consisted of a thorough code inspection aimed
at local improvement. Many malpractices were identified: excessive
use of cut (Section~\ref{section:replace:cut:by:if-then-else}) 
combined with output construction before commit (Section~\ref{section:produce:output:after:commit})
being the most notable one. Additional ``bad smells'' discovered include bad
predicate names such as {\tt q}, unused arguments and unifications
instead of identity checks or numerical equalities (Sections~\ref{section:rename:predicate},
and \ref{section:replace:unification:by:inequality:test}, 
respectively). Some of these were
located by \vipress, others were recognised by the users, while \vipress
performed the corresponding transformations. This step is more
demanding of the user. She has to consider all potential candidates for
refactoring separately and decide on what transformations apply. Hence,
the lion's share of the refactoring time is spent on these local changes.

In summary, from the case study we learned that automatic support for
refactoring techniques is essential and that \vipress is well-suited
for this task. As the result of applying refactoring to \btw we obtained
better-structured lumber-free code. Now it is not only more readable and
understandable but it also simplifies implementing the intended changes. From
our experience with refactoring this large legacy system and the relative
time investments of the global and the local refactorings, we recommend 
starting out with the global ones and then selectively apply local refactorings
as the need occurs. 

The current version of \vipressnospace
can be downloaded from \\
\texttt{http://www.cs.kuleuven.ac.be/\~{}toms/vipress}. 

\section{Conclusions}\label{sec:conclusions}

In this paper we have studied refactoring techniques for Prolog.  Firstly, we
have shown that refactoring is a viable technique for Prolog and that many of
the existing techniques developed for refactoring in general are applicable.
Our refactoring catalogue contains many such refactorings.

Secondly, Prolog-specific refactorings are possible and the application of some
general techniques may be highly specialized towards Prolog. In this context,
the companion technical report \cite{techrep} shows how refactoring fits
in with existing work on program analysis and transformation in the context
of Prolog and how many of these existing techniques may be adapted for the
purpose of partially automating the refactoring process.
Also, \vipressnospace,
our refactoring browser integrates several automatable parts of the presented
refactorings in the VIM editor.

Finally, it should be clear that refactoring Prolog programs is not just viable
but very useful for the maintenance of Prolog programs.  Refactoring helps
bridge the gap between prototypes and real-world applications. Indeed,
extending a prototype to provide additional functionality often leads to
cumbersome code. Refactoring allows software developers both to clean up code
after changes and to prepare code for future changes. These are important
benefits that also apply to logic programming.
 
As completeness of the catalogue is clearly not possible, we aimed to show a
wide range of possibilities for future work on combining the formal techniques
of program analysis and transformation with software engineering. Throughout
the catalogue many specific issues for future work have been mentioned. Below
we list related work and more general challenges for the future.

\subsection{Related and Future Work}


Logic programming has often been used to implement refactorings for other languages, 
e.g. a meta-logic very similar to Prolog is used to detect, for instance,
obsolete parameters in \cite{Tourwe:Mens}.

Seipel {\em et al.}~\shortcite{Seipel:Hopfner:Heumesser}
include refactoring among the analysis and visualization techniques that
can be easily implemented by means of {\sc FnQuery}, a Prolog-inspired query
language for XML. However, the discussion stays at the level of an example.
The M.Sc. thesis of Steinke~\shortcite{Steinke} was dedicated
to refactoring of logic programs. A Catalogue of refactorings has been composed
and a prototype system has been implemented. However, only predicate-scope refactorings
have been considered and only the transformation step has been implemented.   

In the logic programming community questions related to refactoring have been
intensively studied in the context of program transformation and specialisation.
There are two important differences with this line of work. Firstly,
refactoring improves readability, maintainability and extensibility rather
than performance. Secondly,
for refactoring user input is essential while in the mentioned
literature strictly automatic approaches were considered. 
However, some of the transformations developed for
program optimization, e.g. {\em dead code elimination}, can be considered
as refactorings and have an important function in refactoring browsers.

To further increase the level of automation of particular refactorings
additional information such as types and modes can be used. 

Future refactoring tools can also benefit from integration with Prolog development
environments. 
%
Modern Prolog systems are often equipped with features
extending the ISO Standard such as constraint solving over different
domains and Constraint Handling Rules, coroutining, interfaces to foreign
languages, GUI-development systems and databases. In most of the cases,
the refactoring techniques described above can still be applied to
improve the code. 
Certain refactorings may be specially designed for particular
extensions. For instance, our experience suggests that simplifying primitive
constraints may be useful in the case of CLP.


\bibliography{paper}

\bibliographystyle{acmtrans}

\end{document}